\input{aipcheck}

\newcommand{\pom}{\tt I\! P}
\newcommand{\beq}{\begin{equation}}
\newcommand{\eeq}{\end{equation}}

\documentclass[
    ,final            
  ]
  {aipproc}

\layoutstyle{6x9}

\begin{document}

\title{Single and Central Diffractive Higgs Production at the LHC}


\author{M. B. Gay Ducati}{
  address={High Energy Physics Phenomenology Group, GFPAE,  IF-UFRGS \\
Caixa Postal 15051, CEP 91501-970, Porto Alegre, RS, Brazil}
}
\author{M. M. Machado}{
  address={High Energy Physics Phenomenology Group, GFPAE,  IF-UFRGS \\
Caixa Postal 15051, CEP 91501-970, Porto Alegre, RS, Brazil}
}

\author{G. G. Silveira}{
  address={High Energy Physics Phenomenology Group, GFPAE,  IF-UFRGS \\
Caixa Postal 15051, CEP 91501-970, Porto Alegre, RS, Brazil}
}


\begin{abstract}The single and central diffractive production of the Standard Model Higgs boson is computed using the diffractive factorization formalism, taking into account a parametrization for the Pomeron structure function provided by the H1 Collaboration. We compute the cross sections at NLO accuracy for the gluon fusion process, since it is the leading mechanism for the Higgs boson production. The gap survival probability is also introduced to include the rescattering corrections due to spectator particles present in the interaction. The diffractive ratios are predicted for proton-proton collisions at the LHC, since the beam luminosity is favorable to the Higgs boson detection. These results provide updated estimations for the fraction of single and central diffractive events in the LHC kinematical regime.

\end{abstract}

\maketitle


\section{Introduction}

The Higgs mechanism is one of most important subjects to be researched at the LHC, being a cornerstone in the electroweak sector of the Standard Model (SM). We present the calculation of the single diffractive (SD) and the Double Pomeron Exchange (DPE) processes for the production of the SM Higgs boson, considering the diffractive factorization formalism and a recent parametrization for the Pomeron Structure Function provided by the H1 Collaboration \cite{aktas}. The cross sections are computed at NLO accuracy, where we apply to our predictions two different models for the Gap Survival Probability (GSP).

Recent analyses \cite{Tevatron_cut} present an updated mass range for Higgs boson production, which, combining the data coming from CDF and D0 at the Tevatron, have excluded the range 158 < $M_{H}$ < 175 GeV with 95\% of Confidence Level. We estimate the cross sections for the SD and DPE processes and the diffractive ratios for the process $p+p\rightarrow gg \rightarrow H$ in the kinematical regime of the LHC.

\section{Higgs Boson production in $pp$ collisions}

Let us present the main formulae for the Higgs boson production in $pp$ collisions at NLO. The gluon fusion process $pp \rightarrow gg \rightarrow H$ is the leading production mechanism of the Higgs bosons in high-energy $pp$ collisions. The gluon coupling to the Higgs boson in the SM is mediated by a triangular loop of quarks, with a leading contribution from its coupling to the quark top. The partonic cross section at Leading-Order (LO) can be expressed by \cite{spira}
\begin{eqnarray}
\sigma_{LO}(pp\rightarrow H) = \sigma_{0}\tau_{H}\frac{d{\cal{L}}^{gg}}{d\tau_{H}},
\label{sigmaLOlum}
\end{eqnarray}
where the $\sigma_{0}$ reads
\begin{eqnarray}
\sigma_{0} = \frac{G_{F} \alpha^{2}_{s}(\mu^{2})}{288\sqrt{2}\pi}\left| \frac{3}{4}\sum_{q}A_{Q}(\tau_{Q}) \right|^{2},
\label{SIGMAzero}
\end{eqnarray}
where $\mu$ is the renormalization scale, and $A_{Q}(\tau_{Q})=2[\tau_{Q}+(\tau_{Q} +1)f(\tau_{Q})]/\tau_{Q}^{2}$. Considering only the contribution of the quark top, we take $\tau_{Q} = m^{2}_{H}/4m^{2}_{top} \leq 1$, in which $f(\tau_{Q}) = arcsin^{2}{\sqrt{\tau_{Q}}}$ \cite{spira}. The gluon-gluon luminosity is
\begin{eqnarray}
\frac{d{\cal{L}}^{gg}}{d\tau}=\int^{1}_{\tau}\frac{dx}{x}g(x,M^{2})g(\tau/x,M^{2}),
\label{luminosity}
\end{eqnarray}
where $M$ is the factorization scale. The Drell-Yan variable is $\tau_{H}=\frac{m^{2}_{H}}{s}$, with $s$ being the invariant $pp$ collider energy squared.

At Next-to-Leading Order (NLO), the following processes can occur: $gg \rightarrow H$, $gq \rightarrow H$ and $q\bar{q} \rightarrow H$, introducing real and virtual corrections to the $gg \rightarrow H$. The NLO $pp$ cross section for the Higgs boson is written as \cite{spira}
\begin{eqnarray}
\sigma_{NLO}(pp\rightarrow H+X)=\sigma_{0}\left [1+C\frac{\alpha_s(\mu^{2})}{\pi} \right ]\tau_{H}\frac{d{\cal{L}}^{gg}}{d\tau_{H}}+\Delta\sigma_{gg}+\Delta\sigma_{gq}+\Delta\sigma_{q\bar{q}} \, ,
\label{equation38}
\end{eqnarray}
with $\mu$ being the renormalization scale. The expressions for $C(\tau_{Q})$ and for the $\Delta\sigma_{ij}$ can be found in Ref. \cite{spira}, which denotes the virtual and real corrections for the Higgs boson, respectively.

For the single diffractive processes (SD), we consider the Ingelman-Schlein picture \cite{IS}, where the Pomeron structure quark and gluon content is considered. In this approach, the SD cross section is assumed to factorize into the Pomeron-hadron cross section and the Pomeron flux factor. The SD cross section can be written as
\begin{eqnarray}
\sigma^{SD}_{NLO}(M_{H})= C_{g}\int^{1}_{0}\int^{1}_{0} F_{g/p}(x)F_{\pom}(\beta)\sigma_{NLO}(M_{H},\hat{s})d\beta dx
\label{sdexp}
\end{eqnarray}
where $C_{g}$ is a normalization, $\beta=x/x_{\pom}$ and $F_{g/p}(x)$ is the gluon distribution function into the proton. We are considering here the MSTW2008 \cite{mstw} for such a distribution. The Pomeron Structure Function $F_{g/\pom} = f_{\pom/h}(x_{\pom})f_{i/\pom}\left( \frac{x}{x_{\pom}},\mu^{2} \right)$ is expressed by the Pomeron flux $f_{\pom/h}(x_{\pom})$ and the gluon distribution into the Pomeron $f_{i/\pom}(\beta,\mu^{2})$, for which we apply the H1 parametrization \cite{MMM}.

\begin{table}[t]
\centering
\renewcommand{\arraystretch}{0.5}
\begin{tabular}{|c|c|c|c|c|c|c|c|}
\hline
$\rho$  & $\sigma_{\mathrm{inc}}$(pb) & $\sigma_{\mathrm{SD}}$ (pb) & $\sigma_{KKMR}$ (pb)& $\sigma_{GLM}$ (pb) & $R_{Diff} (\%)$ & $R_{KKMR} (\%)$ & $R_{GLM} (\%)$\\ \hline
$0.5$ & $22.01$                     & $0.87$                      & $0.052$             & $0.069$             & $3.95$          & $0.24$          & $0.32$        \\ \hline    
$1.0$ & $16.77$                     & $0.43$                      & $0.026$             & $0.034$             & $2.59$          & $0.16$          & $0.21$        \\ \hline
$1.5$ & $14.45$                     & $0.35$                      & $0.022$             & $0.028$             & $2.48$          & $0.15$          & $0.19$        \\ \hline
$2.0$ & $13.06$                     & $0.32$                      & $0.019$             & $0.026$             & $2.45$          & $0.15$          & $0.19$       \\ \hline
$3.0$ & $11.39$                     & $0.28$                      & $0.017$             & $0.022$             & $2.45$          & $0.15$          & $0.19$        \\ \hline
\end{tabular}  	
\caption{The inclusive cross section ($\sigma_{\mathrm{inc}}$) and the single diffractive cross sections: without absorptive corrections ($\sigma_{\mathrm{SD}}$) and with GSP ($\sigma_{KKMR}$ and $\sigma_{GLM}$), with its respective diffractive ratios $R$, calculated all in NLO (see text).}
\label{tabelaSD}
\end{table}

Moreover, the Double Pomeron Exchange (DPE) cross section reads
\begin{eqnarray}
\sigma^{DPE}_{NLO}(M_{H}) = C_{g}\int^{1}_{0}\int^{1}_{0} F_{\pom_{A}}(x)F_{\pom_{B}}(\beta)\sigma_{NLO}(M_{H},\hat{s})d\beta dx.
\label{sddexp}
\end{eqnarray}
In both diffractive cross sections we introduce the NLO cross section of Eq.(\ref{equation38}).

We further include in Eqs. (\ref{sdexp}) and (\ref{sddexp}) the Gap Survival Probability (GSP) $S_{\mathrm{gap}}^2$ in the cross sections, which can be described in terms of screening or absorptive corrections. It can be estimated using the quantity \cite{Bj}:
 \begin{eqnarray}
<\!|S|^2\!>=\frac{\int|{\cal{A}}\,(s,b)|^2\,e^{-\Omega (s,b)}\,d^2b}{\int|{\cal{A}}\,(s,b)|^2\,d^2b}\,,
 \end{eqnarray}
where $\cal{A}$ is the amplitude, in the impact parameter space, of the particular process of interest at center-of-mass energy $\sqrt{s}$. The quantity $\Omega$ is the opacity (or optical density) of the interaction of the incoming hadrons. This suppression factor of a hard process accompanied by a rapidity gap depends not only on the probability of the initial state survive, but is sensitive to the spatial distribution of partons inside the incoming hadrons, and thus on the dynamics of the whole diffractive part of the scattering matrix. The baseline value in this work are $S_{\mathrm{SD}}^2=6 - 8 \%$ and $S_{\mathrm{DPE}}^2=2.6 - 6 \%$, based on the estimations of the KKMR and GLM models \cite{MMM}.

\section{Results and discussion}

\begin{table}[t]
\centering
\renewcommand{\arraystretch}{0.5}
\begin{tabular}{|c|c|c|c|c|c|c|c|}
\hline
$\rho$  & $\sigma_{\mathrm{inc}}$(pb) & $\sigma_{\mathrm{DPE}}$ (pb) & $\sigma_{KKMR}$ (pb)& $\sigma_{GLM}$ (pb) & $R_{Diff} (\%)$ & $R_{KKMR} (\%)$ & $R_{GLM} (\%)$\\ \hline
$0.5$ & $22.01$                     & $0.12$                       & $0.0033$            & $0.0064$            & $0.57$          & $0.015$         & $0.030$       \\ \hline    
$1.0$ & $16.77$                     & $0.06$                       & $0.0017$            & $0.0040$            & $0.40$          & $0.010$         & $0.024$       \\ \hline
$1.5$ & $14.45$                     & $0.05$                       & $0.0013$            & $0.0031$            & $0.36$          & $0.009$         & $0.022$       \\ \hline
$2.0$ & $13.06$                     & $0.05$                       & $0.0013$            & $0.0031$            & $0.36$          & $0.009$         & $0.022$       \\ \hline
$3.0$ & $11.39$                     & $0.04$                       & $0.0010$            & $0.0024$            & $0.36$          & $0.009$         & $0.022$       \\ \hline
\end{tabular}  	
\caption{The inclusive cross section ($\sigma_{\mathrm{inc}}$) and the diffractive DPE cross sections: without GSP ($\sigma_{\mathrm{DPE}}$) and with GSP ($\sigma_{KKMR}$ and $\sigma_{GLM}$), with its respective diffractive ratios $R$, calculated all in NLO (see text).}
\label{tabelaDPE}
\end{table}

We have calculated the SD and DPE cross sections as well as the diffractive ratios for the LHC energy regime ($\sqrt{s}=14$ TeV) as function of the variable $\rho = \mu/M_{H}$, which shows the relation between $M_{H}$ and the energy scales ($\mu$ = $M$). Our results are presented in Tab.\ref{tabelaSD} for the SD process and Tab.\ref{tabelaDPE} for the DPE process.

It is possible to see that for $\rho \leq$ 2 the estimations to diffractive ratios are more sensitive to $\rho$, since the Higgs mass is similar to the energy scales. However, for larger values of $\rho$, the predictions have a flat behavior. The SD cross section $\sigma_{SD}$ does not consider rescattering corrections, and, in this case, its values are higher than those expected to be observed in the LHC data. However, introducing the rescattering corrections, the SD cross section decreases to values of the order of femtobarns, in agreement to predictions for the exclusive Higgs boson production \cite{KMR,GGS,denterria}. Moreover, our predictions depend on the model taken to estimate the GSP, and one can see that there is a difference by a factor of 2 between the considered models.

These conclusions are more clear in Tab.\ref{tabelaDPE}, where the DPE cross section is evaluated considering the values for the GSP based on the KKMR and GLM models. One observes that the predictions decreases to values of a few femtobarns in both $\sigma_{KKMR}$ and $\sigma_{GLM}$, however with still a difference of about a factor 2 for higher values of $\rho$ and about 3 for lower values.

\section{Conclusions}

In summary, we have evaluated predictions for the diffractive Higgs boson production at the LHC in SD and DPE processes, considering the Ingelman-Schlein picture and introducing rescattering corrections. The cross sections are for both processes can be estimated as $\sigma^{\mathrm{SD}}=  50 - 70$ fb and $\sigma^{\mathrm{DPE}}=  3 - 6$ fb. These cross sections are higher than that obtained from the $\gamma\gamma$ production mechanism, which predict a production cross section of 0.12-0.18 fb \cite{denterria, miller}. Therefore, our estimations are in agreement with previous predictions, however revealing that a detailed study about the GSP in these processes has to be performed.


\begin{theacknowledgments}
This work was supported by CNPq and FAPERGS, Brazil. We want to thanks M. V. T. Machado for useful suggestions and comments.
\end{theacknowledgments}

\end{document}